\documentclass[11pt]{article}

\advance \textheight by \topskip
\usepackage{amsmath,dsfont}
\usepackage[english]{babel}
 \usepackage{color}

\makeatletter
\def\eqnarray{\stepcounter{equation}\let\@currentlabel=\theequation
\global\@eqnswtrue
\global\@eqcnt\z@\tabskip\@centering\let\\=\@eqncr
$$\halign to \displaywidth\bgroup\@eqnsel\hskip\@centering
  $\displaystyle\tabskip\z@{##}$&\global\@eqcnt\@ne
  \hfil$\displaystyle{\hbox{}##\hbox{}}$\hfil
  &\global\@eqcnt\tw@ $\displaystyle\tabskip\z@
  {##}$\hfil\tabskip\@centering&\llap{##}\tabskip\z@\cr}
\@addtoreset{equation}{section}
  \def\theequation{\thesection.\arabic{equation}}
\makeatother

\def\cC{{\mathcal{C}}}
\def\mfQ{{\mathfrak{Q}}}
\def\cH{{\mathcal{H}}}
\def\cK{{\mathcal{K}}}
\def\cD{{{\mathcal{D}}}}

\def\IQ{{\mathds{Q}}}
\def\IS{{\mathds{S}}}

\def\II{{\mathds{1}}}
\def\IO{{0}}

\def\vx{\vec{x}}

\def\vP{\vec{P}}

\newcommand{\p}{\partial}

\newcommand{\cJ}{{\cal J}}

\def\vP{\vec{P}}

\usepackage{amsmath,amssymb}
\def\beq{\begin{equation}}
\def\eeq{\end{equation}}
\def\beqa{\begin{eqnarray}}
\def\eeqa{\end{eqnarray}}
\def\barray{\begin{array}}
\def\earray{\end{array}}

\def\sc{ \scriptscriptstyle \underline }

\begin{document}

\title{
{\bf   Supersymmetry of the planar Dirac --
Deser-Jackiw-Templeton system, and of its non-relativistic
limit }}

\author{
{\sf Peter A. Horv\'athy${}^a$}, {\sf Mikhail S.
Plyushchay${}^{b,c}$}, {\sf Mauricio Valenzuela${}^a$}\,
\footnote{e-mails:
horvathy-at-univ-tours.fr; mplyushc-at-lauca.usach.cl; valenzuela-at-lmpt.univ-tours.fr}\\
[4pt] {\small \it ${}^a$Laboratoire de Math\'ematiques et
de
Physique Th\'eorique, Universit\'e de Tours,}\\
{\small \it Parc de Grandmont,
 F-37200 Tours, France}\\
{\small \it ${}^b$Departamento de F\'{\i}sica, Universidad de
Santiago de Chile, Casilla 307, Santiago 2, Chile}\\
{\small \it ${}^c$Departamento de F\'{\i}sica Te\'orica,
At\'omica y \'Optica, Universidad de Valladolid, 47071,
Valladolid, Spain} }

\date{\today}

\maketitle

\begin{abstract}
The  planar Dirac  and the topologically massive vector
gauge fields are unified into a supermultiplet involving no
auxiliary fields. The superPoincar\'e symmetry emerges from
the $\mathfrak{osp}(1|2)$ supersymmetry realized in terms
of the deformed Heisenberg algebra underlying the
construction. The non-relativistic limit yields spin 1/2
as well as new, spin 1 ``L\'evy-Leblond-type'' equations which,
together, carry an $N=2$ superSchr\"odinger symmetry.
Part of the latter has its origin in the universal
enveloping algebra of the superPoincar\'e algebra.
\end{abstract}

\vskip.5cm\noindent


\section{Introduction}

The two main descriptions of massive relativistic spinning
particles in the plane, namely those with spins $1/2$ and
with spin $1$, are given by the  Dirac, and by the
topologically massive gauge  theory
  of Deser, Jackiw and Templeton  (DJT)~\cite{DJT,Sieg,JT,Schonf},
\beqa
    {\cD}_a^{\ b} \,\psi_b\equiv(P_\mu \gamma^\mu-m)_a^{\ b}\psi_b&=&0\,,
     \label{Dirac}
     \\[4pt]
     \mathfrak{D}_\mu{}^\nu F_\nu\equiv \left(-i\epsilon{}_{
     \mu\lambda}^{\ \ \ \nu} P^\lambda+
    m\delta_\mu^{\ \nu}\right)F_\nu&=&0\,,
\label{DJT}
\eeqa
respectively, where, in \eqref{Dirac},
$\psi=\psi_a$ is a two-component Dirac spinor and the
$-(1/2)\gamma_\mu$, where  $\gamma_0=\sigma^3,\
\gamma_1=i\sigma^2,\ \gamma_2=i\sigma^1$,
 generate the spin $1/2$ representation of the planar Lorentz group.
Similarly in \eqref{DJT},
the $-i\epsilon{}_{
     \mu\lambda}^{\ \ \ \nu}$  generate the $3$-dimensional vector
representation, to which the $F_\nu$ belongs.
The equivalence of the dual formulation we use here with the Chern-Simons approach in \cite{DJT}
 is discussed in \cite{TPN}.

The supersymmetric unification of these two theories is
usually realized in the superfield formulation, which
involves auxiliary fields, and relates the Dirac and the
gauge vector fields. In such an approach, the supersymmetry
between the ``Dirac'' and ``topological'' masses in
(\ref{Dirac}) and (\ref{DJT}), respectively, was noticed in
\cite{DJT, Schonf}, and was extended to topologically
massive supergravity  in \cite{Deser1982}. Below we unify
the two, Dirac and DJT, systems into a single
supermultiplet with no gauge or auxiliary fields.  Our
results differ, hence, from those in \cite{Schonf}.

The non-relativistic limit provides us then with another
minimally realized supersymmetric system composed of the
planar version of L\'evy-Leblond's ``non-relativistic Dirac
equation'' \cite{LLeq}, whose superpartner is a new,
non-relativistic version of the DJT equation we construct
here below.

 Recently \cite{HPV3}, a supersymmetric extension
of Galilei symmetry was obtained by contraction, namely as the
non-relativistic limit of superPoincar\'e symmetry.
Our results here further extend those in \cite{HPV3}~:
the non-relativistic system we
obtain is shown to carry an $N=2$ \emph{superSchr\"odinger symmetry}
\cite{GalSusy,SchrSusy}.

Schr\"odinger symmetry is, in fact, ``\emph{more}'',  and
\emph{not ``less''} than  Poincar\'e symmetry. It is
well-known that the Schr\"odinger symmetry can \emph{not}
be derived  from the relativistic counterpart  by
Inon\"u-Wigner contraction. Extending the usual contraction
to
 the conformal group is, on the one hand, unjustified,
  since the latter is not a symmetry of the (massive)
   relativistic system one starts with. On the other hand,
   the standard contraction procedure
   yields, instead of Schr\"odinger's, the
    \emph{conformal Galilei algebra}, whose physical
    interest is limited  \cite{Barut}.

Here we show, however, that the full superSchr\"odinger symmetry of
the non-relativistic  system emerges by contracting
higher symmetry generators of the relativistic system, namely,
certain elements of the universal enveloping superPoincar\'e algebra.

While the relativistic SUSY has been known before
 \cite{Schonf}, the super-Schr\"odinger symmetry of
 its  non-relativistic counterpart is a new result which,
 to our knowledge, has not been discussed so far.

\section{The Dirac/Deser-Jackiw-Templeton supermultiplet}

We start with considering the direct sum, $D^{1/2}\oplus
D^{1}$,  of the two, spin $1/2$ and spin $1$,
representations, with  Lorentz generators,
\begin{eqnarray}
\cJ_\mu=
\left(
\begin{array}{ccc}
 J^-_\mu & | & 0
 \\
--- & | & ---
\\
 0
 &|&  J^+_\mu
\end{array}
\right) ,\qquad (J^-_\mu)_a{}^b=-\frac{1}{2}
(\gamma_\mu)_a{}^b,\qquad (J^+_\mu)_\nu{}^\lambda
=i\epsilon_{\mu\nu}{}^\lambda\,.
\label{J}
\end{eqnarray}
 A unified wave function can  be represented  by the $5$-tuplet
\begin{equation}
\Psi(x)=
\left(
\begin{array}{c}
\psi_a
\\
--
\\
F_\mu
\end{array}
\right),\quad
(\psi_a)= \left(
\begin{array}{c}
\psi_1(x)  \\
\psi_2(x)
\end{array}\right),
\quad
(F_\mu)= \left(
\begin{array}{c}
F_0(x)  \\
F_1(x)  \\
F_2(x) %
\end{array}
\right).
\label{psi5}
\end{equation}
The 3-dimensional Lorentz algebra generated by $\cJ_\mu$
can be completed
 to the
\emph{superalgebra} $\mathfrak{osp}(1|2)$ by adding the two
off-diagonal matrices 
\begin{equation}
L_{\sc A}=\sqrt{2}
\left(
\begin{array}{ccc}
\IO
 &|& Q_{{\sc A} \,a}{}^\mu\  \\
--- &|& ---\\
 Q_{{\sc A} \,\mu}{}^a &|&  \IO
\end{array}
\right), \label{5a+a-}
\end{equation}
which satisfy $Q_{{\sc A} \,\mu}{}^a= \eta_{\mu\nu} \, (Q_{{\sc A} \,
b}{}^\nu)^T\,\epsilon^{ba}$, where $T$ means 
transposition,  $\epsilon^{12}=-\epsilon^{21}=1,$ and underlined capitals label $\mathfrak{osp}(1|2)$ spinors ($\underline{A}=1,2\,$).
Here the space-time metric is $\eta_{\mu\nu}=diag(-1,1,1)$ and
$\epsilon^{ab}=\epsilon_{ab}$ is used to rise and
lower the spinor indices, $\chi^a=\chi_b\epsilon^{ba}$,
$\chi_a=\epsilon_{ab}\, \chi^b$. Explicitly,
\begin{equation}
Q_{{\sc 1} a}{}^\mu=\left(
\begin{array}{ccc}
 0 & 1 & i \\
 1 & 0 & 0
\end{array}
\right),\quad
Q_{{\sc 2} a}{}^\mu=\left(
\begin{array}{ccc}
 1 & 0 & 0 \\
 0 & 1 & -i
\end{array}\right),
\end{equation}
related as $(Q_{{\sc 1}a}{}^\mu)^\dagger = -Q_{{\sc
2}\mu}{}^a\,,\, (Q_{{\sc 1}\mu}{}^a)^\dagger = Q_{{\sc
2}a}{}^\mu$.  The operators $L_{\sc 1}$, $L_{\sc 2}$ are
hermitian conjugate with respect to the scalar product,
$\Phi^\dagger \hat{\eta} \Psi$. Hence $L_{\sc 1}=\hat{\eta}
L_{\sc 2}^\dagger \hat{\eta}$, where
$\hat{\eta}=diag(\gamma^0,\eta)$, and $\eta=
\eta_{\mu\nu}$.

$L_{\sc 1}$ and $L_{\sc 2}$ span the deformed Heisenberg
algebra \cite{DHA}, $ \big[L_{\sc A}\,,L_{\sc
B}\big]\!=-\epsilon_{{\sc A}{\sc B}}\,(1+\nu{}R)\,, $ with
deformation parameter $\nu=-5$, where
\beq\label{RZ2}
R={\rm diag}\big(-\II_2,\II_3\big),
\eeq
 $R^2~=~\II,\, \{L_{\sc A},R\}=0\,,$ is the reflection operator.
The operators $L_{\sc A}$ extend the Lorentz algebra generated by \eqref{J}
 to $\mathfrak{osp}(1|2)$,
\begin{equation}\label{osp12}
[{\cal J}_\mu, {\cal J}_\nu]=-i\epsilon_{\mu\nu\lambda} {\cal J}^\lambda\,, \qquad
\{L_{\sc A},L_{\sc B}\}=4 ({\cal J}\gamma)_{{\sc A}{\sc B}}\,,\qquad
[{\cal J}_{\mu},L_{\sc A}]= \frac{1}{2}(\gamma_{\mu})_{\sc A}{}^{\sc B}L_{\sc B}\,,
\end{equation}
where $(\gamma^\mu)_{{\sc A}{\sc B}}= \epsilon_{{\sc B}{\sc
C}}(\gamma^\mu)_{\sc A}{}^{\sc C}\,$. The role
of the grading operator is played by the reflection operator $R$.
 Then  the super-Casimir operator is
\begin{equation}
     \mathcal{ C}=\cJ_\mu \cJ^\mu+\frac{1}{8}
     [L_{\sc 1},L_{\sc 2}]=\cJ_\mu \cJ^\mu+\frac{1}{8}
     L^{\sc A} L_{\sc A}=-\frac{3}{2}\ .
\end{equation}
The representation of $\mathfrak{osp}(1|2)$  is therefore
irreducible. The original ingredients, $\psi_a$ and
$F_\mu$, can plainly be recovered by projecting onto the $\mp 1$  eigenspaces of  the
reflection operator, $R$.
 On these subspaces the Casimir of the Lorentz subalgebra is
\begin{equation}
\label{JJ}
    \cJ_\mu \cJ^\mu=-\hat{\alpha}(\hat{\alpha}-1)\,
\quad\hbox{with}\quad
    \hat{\alpha}=-\frac{1}{4}(3+R)\,.
\end{equation}
The operator $\hat{\alpha}$  has, hence, eigenvalues
$ \alpha_-=-\frac{1}{2}\,$
 and
$\alpha_+=-1\,$,
proving that the $\mp 1$ eigenspaces carry indeed
the irreducible spin-$1/2$ (Dirac) and resp. spin-$1$ DJT representations.
Moreover, using $\hat{\alpha}$ our two, Dirac and DJT,
systems can be written in the same unified form,
\begin{equation}
(P_{\mu}\cJ^\mu-\hat{\alpha} m)\Psi=0.
\label{SM}
\end{equation}

The operators $L_{\sc A}$ interchange $\psi$ and $F$, but
they do not preserve the physical states (defined as
solutions of the Dirac and DJT equations, respectively).
This can be achieved, however, by considering instead the
two supercharges,  
$$\mathcal{Q}_{\sc A}=\frac{1}{2\sqrt{m}}(P_\mu (\gamma^\mu)_{\sc A}{}^{\sc B}- R m \,\delta_{\sc A}{}^{\sc B})\, L_{\sc B},$$
 $A,B=1,2$, whose components in explicit form are, 
\begin{equation}\label{Q}
\mathcal{Q}_{\sc 1}=\frac{1}{2\sqrt{m}}\left(
  L_{\sc 2} P_+ + L_{\sc 1} (m R- P_0)\right),
\qquad
\mathcal{Q}_{\sc 2}=\frac{1}{2\sqrt{m}}\left(
- L_{\sc 1} P_- + L_{\sc 2} (mR+ P_0)\right),
\end{equation}
%

where $P_\pm=P_1\pm iP_2$. The action of (\ref{Q}) on the spin-1
($F_\mu$) and spin-1/2 ($\psi_a$) components
 is found, respectively, as
\begin{eqnarray}
 \Psi'=\left(\begin{array}{c}
\psi'_a
\\
F'_\mu
\end{array}\right)
= \zeta^{\sc A}\mathcal{Q}_{\sc A}\Psi= \zeta^{\sc A}\left(
\begin{array}{c}
\mathcal{Q}_{{\sc A} a}{}^\mu F_\mu
\\
\mathcal{Q}_{{\sc A} \mu}{}^a \psi_a
\end{array}
\right), \label{psi'F'}
\end{eqnarray}
where $\zeta^{\sc A}$ are the parameters of the
supersymmetry transformation. Hence, a two-component Dirac
field is transformed  into a three-component DJT field $F'$
and conversely. Furthermore,
\begin{eqnarray}
    \mathcal{D}_a{}^b \,\psi'_b&=&\zeta^{\sc A}\left(
    \mathcal{Q}_{{\sc A} a}{}^\mu \mathfrak{D}_\mu{}^\nu
    F_\nu +\frac{1}{2\sqrt{m}}
    Q_{{\sc A} a}{}^\mu (P^2+m^2) F_\mu\right) \,,
    \label{Dpsi'}
    \label{susytr2a}
    \\[4pt]
    \mathfrak{D}_\mu{}^\nu F'_\nu&=&\zeta^{\sc A}\left(-
    \frac{1}{2} \mathcal{Q}_{{\sc A} \mu}{}^a
    \mathcal{D}_a{}^b \psi_b - \frac{1}{2\sqrt{m}}
    Q_{{\sc A} \mu}{}^a  (P^2+m^2) \psi_a\right)\,.
    \label{DF'}
    \label{susytr2b}
\end{eqnarray}
 Both the Dirac and DJT equations imply the Klein-Gordon equation,
  allowing us to conclude that
the transformed fields satisfy the Dirac and DJT equations,
respectively, if the original ones satisfy them (in the
reversed order).

Adding the two supercharges  (\ref{Q})  to the Poincar\'e
generators of the space-time translations, $P_\mu$, and of
the Lorentz transformations,
$\mathcal{M}_\mu=-\epsilon_{\mu\nu\lambda}x^\nu
    P^\lambda+\cJ_\mu$,  yields the off-shell relations,
\begin{equation}
 [P_\mu,P_\nu]=0\,,\qquad
    [{\cal M}_\mu, P_\nu]=-i\epsilon_{\mu\nu\lambda}P^\lambda\,,\qquad
    [{\cal M}_\mu,{\cal M}_\nu]=-i\epsilon_{\mu\nu\lambda}{\cal
    M}^\lambda ,\label{Poinc}
\end{equation}
\begin{equation}
[P_\mu,\mathcal{Q}_{\sc A} ]=0 \,,\qquad [{\cal M}_\mu,\mathcal{Q}_{\sc
A}]=\frac{1}{2}(\gamma_\mu)_{\sc A}{}^{\sc B}
   \mathcal{Q}_{\sc B}\,,
\label{SuperPoi1}
\end{equation}
\begin{eqnarray}
   \{\mathcal{Q}_{\sc A},\mathcal{Q}_{\sc B}\} &=&2(P\gamma)_{{\sc A}{\sc
B}}\label{QQPab}
    +\displaystyle\frac{1}{2m}\left[({\cal J}\gamma)_{{\sc A}{\sc
B}}(P^2+m^2)-
    2(P\gamma)_{{\sc A}{\sc B}}(P{\cal J}-
    \hat{\alpha}m)\right]\,.\label{SuperPoi3}
\end{eqnarray}
The second term  on the r.h.s. of \eqref{SuperPoi3}
vanishes on-shell, leaving us with the usual $N=1$ planar
super-Poincar\'e  algebra, $\mathfrak{iso}(1|2,1)$.

\section{Solutions of the supersymmetric equation}

The equations are solved following the method outlined in
\cite{HPV3}. It  requires expanding  the fields in the
lowest weight representation basis of the Lorentz
generators \eqref{J}.

Since the representation of the Lorentz algebra \eqref{J} is
reducible, there are two lowest nontrivial vectors such
that, $\mathcal{J}_-|0)_D=0$, $\mathcal{J}_-|0)_{DJT}=0$.
These are just the lowest spin states in the spin $1/2$
(Dirac) and spin $1$ (DJT) sectors, respectively.

The irreducible spaces of spin $1/2$ and $1$ representations are
generated by the ladder operator
$\mathcal{J}_+=\mathcal{J}_1+i\mathcal{J}_2$,  which  acts as
\beq
|1)_D=\mathcal{J}_+|0)_D\,,\quad
|1)_{DJT}=\frac{1}{\sqrt{2}}\mathcal{J}_+|0)_{DJT}\,,\quad
|2)_{DJT}=\frac{1}{\sqrt{2}}\mathcal{J}_+|1)_{DJT}\,.
\eeq
Both subspaces have highest spin states,
$\mathcal{J}_+|1)_D=0$, $\mathcal{J}_+|2)_{DJT}=0$ and the
ladder operator $J_-=\mathcal{J}_1-i\mathcal{J}_2$ acts as,
\beq
\mathcal{J}_-|1)_D=-|0)_D\,, \quad
\frac{1}{\sqrt{2}}\mathcal{J}_-|1)_{DJT}=-|0)_{DJT}\,,
\quad
\frac{1}{\sqrt{2}}\mathcal{J}_-|2)_{DJT}=-|1)_{DJT}\,.
\eeq
$\mathcal{J}_0$ acts diagonally,
\beqa
\mathcal{J}_0|0)_D=-\frac{1}{2}|0)_D,\qquad
\mathcal{J}_0|1)_D=\frac{1}{2}|0)_D,
\\[6pt]
 \mathcal{J}_0|0)_{DJT}=-|0)_{DJT}, \qquad
 \mathcal{J}_0|1)_{DJT}=0,\qquad
 \mathcal{J}_0|2)_{DJT}=|2)_{DJT}\,.
\eeqa
The Dirac and DJT fields  are written in this basis as,
\beq
\psi= \psi_0 |0)_D + \psi_1 |1)_D\,, \quad
F=\frac{1}{\sqrt{2}}\,F_+ |0)_{DJT} + F_0 |1)_{DJT} +
\frac{1}{\sqrt{2}}\,F_- |2)_{DJT}\, .
\eeq
Here, $F_\pm=F_1\pm iF_2$.

For the Dirac and DJT fields,
equation \eqref{SM} yields,
\begin{equation}
    \frac{1}{\sqrt{2}}\left[(m-P^0)\psi_0 + P_+\psi_1
    \right]|0)_D+\frac{1}{\sqrt{2}}\left[(m+P^0)\psi_1 +
     P_-\psi_0 \right]|1)_D=0, \label{Dirc}
\end{equation}
\begin{eqnarray}
    \frac{1}{\sqrt{2}}\Big[(m+P_0)F_+ - P_+ F_0
    \Big]|0)_{DJT}  +
     \Big[m F_0 + \frac{P_-F_+-P_+F_-}{2} \big]|1)_{DJT}\nonumber
 \\[4pt]
 +\frac{1}{\sqrt{2}}\Big[(m-P_0)F_0
    + P_- F_0 \Big]|2)_{DJT}=0\,.
     \label{DJTc}
\end{eqnarray}
For positive energy solutions ($P^0=-P_0>0$), the operator $P^0+m$ can be inverted. Hence,
\begin{eqnarray}
     \frac{(P\cdot P+m^2)}{(P^0+m)^2}F_0+F_0  =
     -\frac{P_-}{P^0+m} F_+, \quad F_- =
     \left(\frac{P_-}{P^0+m}\right)^2 F_+,\quad \psi_1
     =-\frac{P_-}{P^0+m} \psi_0 \,.\label{sol}
\end{eqnarray}
The first term in \eqref{sol} vanishes by the Klein-Gordon
equation, so that all components of the  DJT (Dirac) field
are obtained from the lowest spin state $-1$ (and $-1/2$),
\begin{eqnarray}
    F_+ = A\, \varPhi(x),\quad  \psi_0 =B\,\varPhi(x)\,, \qquad
     \varPhi(x)=\exp \left\{ -ix_0\sqrt{p_i^2+m^2}+ix_ip_i
    \right\},
\end{eqnarray}
where $A$, $B$, are arbitrary constants,
and
$p_i$ are the eigenvalues of $P_i$.

Negative energy solutions can be obtained by an analogous procedure,
considering the highest spin components,
$\psi_2$ and $F_-$, of the spin $1/2$ and resp
$1$ sectors.

\section{Nonrelativistic counterpart of the Dirac-DJT supermultiplet}
\label{D-DJTSUSY}

Taking the nonrelativistic limit is subtle. For example,
central extensions correspond to cohomology
\cite{cohom,SSD,AldAzc,cohom+}; that of the Poincar\'e
group is trivial, while the one of the Galilei group is
not. How can nontrivial cohomology arise in the
nonrelativistic limit ? As explained in Ref. \cite{AldAzc},
one should start with the trivial $U(1)$ extension (i.e.
with trivial two-cocycle) of the universal covering of the
Poincar\'e group, and then In\"on\"u-Wigner contraction
yields the universal covering of the Galilei group, namely
an $U(1)$ extension of the Galilei group (with nontrivial
two-cocycle). The latter is necessary to support the
mass-central-charge extension. It is in fact the rest frame
energy $mc^2$ that generates the nontrivial two-cocycle in
the nonrelativistic limit.

In our particular case, the nonrelativistic limit is
carried out first by reinstating the velocity of light,
$c$, and putting $m\rightarrow mc$, $x^0=ct$. $P^0$
diverges in the nonrelativistic limit $c\rightarrow \infty$ as
$mc$ and must be renormalized therefore.  Similar
considerations indicate that, when compared to $F_+$, the
components $F_0$ and $F_-$ are suppressed by factors of
order $c^{-1}$ and $c^{-2}$, respectively, on account of
equation \eqref{sol}. Analogously for the Dirac field,
$\psi_1$ is suppressed by $c^{-1}$ compared $\psi_0$.
 Rescaling the field
components by suitable powers of $c$ yields nontrivial
components  with nonrelativistic spin. Consider, in fact, $
{\phi}_0= e^{-imc^2t} \psi_0\,,\, {\phi}_1=c e^{-imc^2t}
\psi_1 $ for the Dirac field, and $ f_+= e^{-imc^2t}
F_+\,,\, f_0=c e^{-imc^2t} F_0\,, \, f_-=c^2 e^{-imc^2t}
F_- $ for the DJT field. These transformations can be
written in a compact form in terms of the supermultiplet
\eqref{psi5},  namely as $ \mathbf{\Phi}=\mathbf{M}\,
\Psi,\; \mathbf{M}=diag(\mathbf{M}^-,\mathbf{M}^+) \,, $
where $\mathbf{M}$  is a block-diagonal matrix, composed of
\begin{equation}
 \mathbf{M}^- = e^{-imc^2t}  diag(1,\,c),\qquad \mathbf{M}^+= e^{-imc^2t} \left(
\begin{array}{ccc}
 c & 0 & 0 \\
 0 & 1+c^2 & i(1-c^2) \\
 0 & -i(1-c^2) & 1+c^2
\end{array}
\right).
\end{equation}
The relativistic operators transform according to
\beq
{\cal
O}\to{\cal O}'=\mathbf{M} {\cal O}\mathbf{M}^{-1}.
\label{similarity}
\eeq
In terms of
\begin{equation}
P'_0= -c^{-1}(i\p_t+mc^2),\quad P'_i=P_i,\quad
\cJ'_0=\cJ_0,\quad \cJ'_+=c\cJ_+\,,\quad \cJ'_-=\frac{1}{c}
\cJ_-\,  \label{Pprime}
\end{equation}
Eqn \eqref{SM} can therefore be rewritten as $
(P'_{\mu}\cJ'^\mu-\hat{\alpha} mc)\mathbf{\Phi}=0 $.
Switching to primed variables is, in fact, an authomorphism
of the Poincar\'e algebra \eqref{Poinc}, so that the value
of the $\mathfrak{so}(2,1)$ Casimir operator of
 $\cJ'$ is left invariant. We have, moreover,
\begin{equation}
P'_{\mu}\cJ'^\mu-\hat{\alpha} mc = \frac{1}{c}
\left(i\cJ_0\partial_t+ \frac{1}{2} P_+\cJ_-\right)+c
\left(m(\cJ_0-\hat{\alpha})+ \frac{1}{2} P_-\cJ_+\right).
\label{NRprime}
\end{equation}
This operator diverges in the nonrelativistic limit. Consistency
 in the nonrelativistic limit
 requires, therefore,
\begin{equation}
\left(m(\cJ_0-\hat{\alpha})+ \frac{1}{2}
P_-\cJ_+\right) \mathbf{\Phi}=0. \label{eqNR1}
\end{equation}
In the rest frame, this equation is equivalent to
$\cJ_0-\hat{\alpha}=0$, which fixes the spin of the
nonrelativistic particle.  Note that the nonrelativistic
Hamiltonian, ${\cH}=i\partial_t=cP^0-mc^2$, is \emph{not}
 obtained here, since the first term in (\ref{NRprime})
drops out when $c\to\infty$. The Schr\"odinger equation is
obtained, however,  from the transformed Klein-Gordon
equation [which is, as
 said before, a consequence of the first-order equations (\ref{Dirac}) and
 (\ref{DJT})],
\begin{equation}
(i\partial_t -\frac{1}{2m} P_+ P_-)\mathbf{\Phi} =
\lim_{c\to\infty}\left(-\frac{1}{2m}  (P'^2+m^2c^2)\mathbf{\Phi}\right)=0.
\label{KGSch}
\end{equation}
Furthermore, \eqref{eqNR1} and \eqref{KGSch} allow us to infer,
\begin{eqnarray}
&\left((\cJ_0-\hat{\alpha})P_+ +  i\cJ_+ \, \partial_t \right) \mathbf{\Phi}=0\,.&
\label{eqNR2}
\end{eqnarray}
Eqn (\ref{eqNR2}),  together with  \eqref{eqNR1} allows us
to recover, once again, the Schr\"odinger equation
$(i\partial_t-(2m)^{-1} P_i^2)\mathbf{\Phi}=0$ as
consistency condition, namely
 by commuting the operators in front the field $\mathbf{\Phi}$.
Hence,
\eqref{eqNR1}, \eqref{KGSch} and \eqref{eqNR2} form a self consistent system.
In fact, Eqns \eqref{eqNR1} and \eqref{eqNR2} alone are enough
 to describe our massive nonrelativistic supermultiplet.
Projecting these equations to the spin $1/2$ and $1$
subspaces yields indeed the independent equations (written in component form),
\begin{eqnarray}
\hbox{spin } \frac{1}{2} \quad &:&\qquad
    \left\{\barray{lll}
    i\partial_t
    {\phi}_0 + P_+{\phi}_1&=&0\,,
     \\[6pt]
      2m{\phi}_1 + P_-{\phi}_0 &=&0\,,
      \earray\right.
\label{LLequation}
\\[12pt]
\hbox{spin } 1 \quad &:& \qquad
\left\{\begin{array}{lll}
i \partial_t f_+ -i
    P_+ f_0&=&0\, ,
    \\[6pt]
2mf_0 + i P_-f_+ &=&0 \,,
\\[6pt]
  2m f_- + iP_-f_0 &=&0
\,.
\end{array}\right.
\label{NRDJT}
\end{eqnarray}
Equations \eqref{LLequation} are
the $(2+1)$D L\'evy-Leblond equations \cite{LLeq}.  (\ref{NRDJT}) is in turn the
non-relativistic limit of the spin-$1$ Deser-Jackiw-Templeton system, cf. \cite{HPV3}.
$\psi_1$, $f_0$ and $f_-$ are auxiliary
fields and may be expressed in terms of the lowest spin states,  $\psi_0$ and $f_+$, respectively.

These equations imply the Schr\"odinger equation for each component.

With the nonrelativistic limit  is associated a contraction
of the superPoincar\'e algebra \eqref{Poinc}, that produces
a symmetry of the nonrelativistic system \eqref{KGSch} and
\eqref{eqNR2}. Defining $
{\cK}_i=-\lim_{c\rightarrow\infty}\epsilon_{ij}\mathcal{M}'_j/c\,,
\
    {\cH}=cP'^0-mc^2\,, \ {\rm J}=\mathcal{M}'_0\,
$
where $\epsilon_{ij}=-\epsilon_{ji}$, $\epsilon_{12}=1$, we get
\begin{equation}
    \label{KJM1}
   {\cK}_i=-tP_i+m x_i+\mathfrak{K}_i\,, \qquad
    {\cH}=i\frac{\partial}{\partial t}\,, \qquad
    {\rm J}=\epsilon_{ij}x_iP_j + \mathcal{J}_0\,,
\end{equation}
where $\mathfrak{K}_1=i\mathfrak{K}_2=\frac{i}{2}\mathcal{J}_+$.
 $\mathcal{J}_+$  has diagonal blocks $\mathcal{J}_+=
 diag(\mathcal{J}_+\Pi_-,\mathcal{J}_+\Pi_+)$
 which act independently in the spin $1/2$  and spin $1$ subspaces,
\begin{equation}
\mathcal{J}_+\Pi_-\,=
\left(
\begin{array}{cc}
0 & 0 \\[4pt]
1 & 0
\end{array}
\right),
\qquad
 \mathcal{J}_+\Pi_+\,=
\left(
\begin{array}{ccc}
0 & 1 & i \\[4pt]
1 & 0 & 0 \\[4pt]
i & 0 & 0
\end{array}
\right),\label{frakK}
\end{equation}
where $\Pi_\pm=\frac{1}{2}(1\pm R)$. The boost operator we
find is consistent with the known result for spin 1/2
\cite{LLeq}, recently generalized to  spin 1  \cite{HPV3}.
Note the spin contribution, $\mathcal{J}_0$, to the angular
momentum. Together with the $P_i$, the operators
(\ref{KJM1}) generate the seven-dimensional,
 one-parameter centrally extended
Galilei (also called ``Bargmann'') algebra
$
\mathfrak{gal}=
\Big\{\hbox{translations}, P_i,\,
\hbox{time translations}, {\cH},\,
\hbox{Galilei boosts}, {\cK}_i,\,
\hbox{rotations}, {\rm J},\,
\hbox{mass-central-charge}, m
\Big\},
$
\begin{eqnarray}
\begin{array}{c}
    [{\cal K}_i,P_j]=im\delta_{ij}\,,\quad [P_i,P_j]=0\,,
    \quad [{\cH},P_i]=0\,, \quad
    [{\cK}_i,{\cK}_j]=0\,, \\[8pt]
    [{\cK}_i,{\cH}]=iP_i\,,\quad
    [{\rm J},P_i]=i\epsilon_{ij}P_j\,,\quad
    [{\rm J},{\cK}_i]=i\epsilon_{ij}
    {\cK}_j\,.
    \label{Gal}
\end{array}
\end{eqnarray}
The Casimir operators of the algebra (\ref{Gal}) are
\begin{equation}\label{Cas}
 \mathcal{C}_1=P_i^2-2m{\cal H}\,,\qquad
    \mathcal{C}_2=m{\rm J}-\epsilon_{ij}{\cK}_i P_j=
    m\mathcal{J}_0 + \frac{1}{2} P_-\mathcal{J}_+ \,.
\end{equation}
The internal energy represented by  $\mathcal{C}_1$ vanishes, owing to the Schr\"odinger
equation. In virtue of equation \eqref{eqNR1}, the second Casimir is, however, operator valued,
$$
\mathcal{C}_2=m \hat{\alpha}\,.
$$
The Galilei algebra we have found is therefore reducible.
This has been expected, owing to the supersymmetry of the Dirac-DJT multiplet,
whose nonrelativistic limit we have taken. The algebra becomes, however,
irreducible if we restrict ourselves to the $\mp1$ subspaces of the operator
$R$, i.e., to spin $1/2$ and spin $1$, respectively.

Non-relativistic supersymmetry is inherited from the relativistic
one of the Dirac-DJT supermultiplet.
In fact, the nonrelativistic supercharges, are obtained by taking the limit,
${\mfQ}_{\sc A} = \lim_{c\rightarrow \infty} \frac{1}{c}\mathcal{Q}'_{\sc A}\,$,
of transformed expressions
$\mathcal{Q}'_{\sc A}=\mathbf{M}\mathcal{Q}_{\sc A}\mathbf{M}^{-1}$.
 Explicitly, for
 $\mfQ_1=\frac{1}{\sqrt{2}}(\mfQ_{\sc 1}+\mfQ_{\sc 2})$ and
 $\mfQ_2= \frac{1}{\sqrt{2}\,i}(\mfQ_{\sc 1}-\mfQ_{\sc 2})$ we find
\beqa
{\mfQ}_1&=&\frac{1}{\sqrt{2}}
\left(
\begin{array}{ccc}
0
 &|& \sqrt{m}\,Q_{{\sc 1} \,a}{}^\mu\
\\
---------- &|& ---\\
-\frac{1}{2\sqrt{m}}\left(P_-Q_{{\sc 1} \,\mu}{}^a +
 2m P_-Q_{{\sc 2} \,\mu}{}^a \right) &|&  0
\end{array}
\right),\label{Q1NR}
\\[8pt]
{\mfQ}_2&=&\frac{1}{\sqrt{2}i}
\left(
\begin{array}{ccc}
0
 &|& \sqrt{m}\,Q_{{\sc 1} \,a}{}^\mu\
\\
---------- &|& ---\\
\frac{1}{2\sqrt{m}}\left(P_-Q_{{\sc 1} \,\mu}{}^a +
2m P_-Q_{{\sc 2} \,\mu}{}^a \right) &|&  0
\end{array}
\right). \label{Q2NR} \eeqa
Note that these
supercharges are related through the
grading operator (\ref{RZ2}), ${\mfQ}_2=iR\,{\mfQ}_1$.
 The action of the supercharge
${\mfQ}_1$
    on the nonrelativistic field $\bf{\Phi}$ reads
\begin{eqnarray}
 \bf{\Phi}'=
\left(\begin{array}{c}
\phi'_a
\\[10pt]
f'_\mu
\end{array}\right)
=
\left(
\begin{array}{c}
\sqrt{m}Q_{{\sc 1} a}{}^\mu f_\mu
\\[10pt]
-\frac{1}{2\sqrt{m}}\left(P_-Q_{{\sc 1}
\,\mu}{}^a + 2m P_-Q_{{\sc 2} \,\mu}{}^a \right)\phi_a
\end{array}
\right),
\label{psi'f'}
\end{eqnarray}
and the action of  ${\mfQ}_2$ follows analogously.
Here, $f_1 = (f_+ + f_-)/2$ and, $f_2 =-i(f_+ - f_-)/2.$
This formula shows how the supercharges indeed interchange the non-relativistic
``Dirac'' ($\phi_a$) and ``DJT'' ($f_\mu$)
components, cf. (\ref{psi'F'}).  Our supercharges
$\mfQ_i$,  extend  \eqref{Gal} to a superalgebra  \cite{GalSusy} with (anti)commutation relations,
\begin{eqnarray}
\begin{array}{c}
[{\rm J},{\mfQ}_i]=i\epsilon_{ij}\mfQ_j \,,
 \qquad
\{{\mfQ}_i,{\mfQ}_j\}=\delta_{ij}\left( 4m +
\left(m(\cJ_0-\hat{\alpha})+ \frac{1}{2} P_-\cJ_+\right)\right)\,,
     \\[12pt]
[{\cK}_i,{\mfQ}_i]=
[P_i,{\mfQ}_i]=[{\cH},\mathfrak{
    Q}_i]=0\,.
\end{array} \label{GalQ}
\end{eqnarray}
Note that  the supercharge is a vector w.r.t. a rotation.
Observe also that the supercharge-anticommutator involves
the operator in Eqn. \eqref{eqNR1}, which vanishes when
acting on a wave function on account of  Eqn.
(\ref{eqNR1}). On-shell we have therefore
\begin{equation}\label{mcentral}
    \{{\mfQ}_i,{\mfQ}_j\}=4\delta_{ij}\,m.
\end{equation}

The  non-relativistic theory  actually has \emph{more
symmetries}, which do \emph{not} derive directly from the
Lie (super)algebraic structure of the parent relativistic
theory \cite{GalSusy,SchrSusy} (see the discussion below).
The two ``helicities''
 \beq
    \IQ=\frac{1}{2m}\vec{\mfQ}\cdot\vP, \qquad
    \IQ^\star=\frac{1}{2m}\vec{\mfQ}\times\vP\,,
    \label{helicity}
\eeq
related by $\IQ^\star=iR\IQ$,
 are both new supercharges for the
non-relativistic system which
yield on-shell, with the previous operators, a
 closed superalgebra,
\begin{eqnarray}
\begin{array}{llll}
[\IQ,\,\mathcal{K}_i ]=-\frac{i}{2}\mfQ_i \,,& [\IQ,\,P_i ]=0 \,, & [\IQ,\,\cH]=0\,,
\\[6pt]
[\IQ^\star,\,\mathcal{K}_i ]=\frac{i}{2}\epsilon_{ij}\mfQ_j \,,
& [\IQ^\star,\,P_i ]=0 \,, & [\IQ^\star,\,\mathcal{H} ]=0\,,
\\[6pt]
\{ \IQ, \, \IQ \}=2\cH,
&\{ \IQ^\star, \, \IQ^\star \}=2\cH,
&\{ \IQ, \, \IQ^\star \}=0\,,
\\[6pt]
\{\mfQ_i,\,\IQ \}=2P_i,
&\{\mfQ_i,\,\IQ^\star \}=2\epsilon_{ij} P_j,
&
\end{array}
\end{eqnarray}
called the [centrally extended] $N=2$ superGalilei algebra
\cite{GalSusy,SchrSusy}, $\mathfrak{sgal}$, which extends
the Galilei algebra, $\mathfrak{gal}$, by the odd
supercharges $\mfQ_i$, $\IQ$, and $\IQ^\star $. In
particular, \emph{both} $\IQ$ and $\IQ^\star$ are ``square
roots'' of the Hamiltonian, ${\cH}$ -- just like the
${\mfQ}_i$ in (\ref{mcentral}) are ``square roots'' of the
mass. The super-Casimir operator reads \beq
\mathcal{C}_{susy}=\mathcal{C}_{2}+\frac{i}{16}[{\mfQ}_{1},\,{\mfQ}_{2}]=-\frac{3}{4}m\,,
\eeq where \eqref{eqNR1} was taken into account. Note that
the representation of the superalgebra is irreducible,
since $\mathcal{C}_{susy}$ is a constant.

By (\ref{helicity}), and remembering that
$m$ plays the role of the central charge, our
$N=2$ superGalilei algebra  also has  two odd Casimir operators, namely
\begin{equation}\label{c3c4}
    \mathcal{C}_3=m\IQ-\frac{1}{2}\vec{\mfQ}\cdot\vP\,, \qquad
    \mathcal{C}_4=m\IQ^*-\frac{1}{2}\vec{\mfQ}\times\vP\,.
\end{equation}
They take here zero values.

\section{Schr\"odinger (super)symmetry}

For a spinless particle,
 the (free) Schr\"odinger equation is known to be symmetric under the ``conformal''
 extension of the Galilei group, obtained by
 adding dilations  and expansions  \cite{SchrSymm},
\beq D=2t\cH-\vx\cdot \vP+i\,, \qquad C=t^2\cH-t \vx\cdot
\vP+it+\frac{m}{2}\vx{\strut}{\,}^2. \label{spin0Sch} \eeq

 Since the nonrelativistic spin $1/2$ and spin $1$ equations,
 \eqref{LLequation} and \eqref{NRDJT},
 also  describe free particles, their Schr\"odinger symmetry is expected (and has
actually been proved for the spin $1/2$ model of
L\'evy-Leblond \cite{DHP}.)  Now we prove that  the operators
\begin{eqnarray}
\begin{array}{ll}
{\cD}=-\displaystyle\frac{1}{2m} ({\cK}\cdot P + P \cdot {\cK})=
D\;- \displaystyle\frac{i}{2m} P_- \mathcal{J}_+,\\
[8pt]
\mathcal{C}=\displaystyle\frac{1}{2m} {\cK}\cdot
{\cK}=C+ \displaystyle\frac{i}{2m}(-tP_-+mx_-)\, \mathcal{J}_+,
\end{array}\label{DC}
\end{eqnarray}
extend the Galilei algebra \eqref{Gal} into
 the Schr\"odinger algebra,
$\mathfrak{sch}$, with non-trivial additional commutation
relations,
\begin{equation}
\begin{array}{c}
[{\cD},\,\mathcal{C}]=2i\mathcal{C},
\qquad
[{\cD},\,{\cH}]=-2i{\cH},
\qquad
[{\cH},\,\mathcal{C}]=i{\cD},
\\[6pt]
[{\cD},\, {P}_i] = -i {P}_i, \qquad [D,\,\cK_i] = i\cK_i,\qquad
[\mathcal{C},\, P_i] =  i{\cK}_i\,.
\end{array}
\label{CDH}
\end{equation}
 This representation is reducible:  the operators  in (\ref{DC})
act diagonally on the spin $1/2$ and spin $1$ subsystems.
Projected to the spin-$1/2$ subspace we obtain, using
$P_-\mathcal{J}_+=-2m(\mathcal{J}_0-\hat{\alpha})$ (cf. Eqn.
\eqref{eqNR1}), \begin{equation} {\cD}^- = {\cD}\,
\Pi_-\approx\left(
\begin{array}{ll}
D & 0 \\
0 & D+i
\end{array}\right)\,
,\quad
\mathcal{C}^- = \mathcal{C}\,\Pi_- \approx \left(
\begin{array}{ll}
C & 0 \\
\frac{i}{2}x_- &C+it
\end{array}\right)\,,
\end{equation}
which are equivalent, on-shell, with those  in Ref. \cite{DHP}.

Projecting instead onto the spin-$1$ sector we obtain the new
result,
\begin{equation}
{\cD}^+ = D\, \Pi_+\approx \left(
\begin{array}{lll}
D+i & 0 & 0 \\
0 &D+i & 1\\
0 & -1 &D+i
\end{array}\right)\,
,\quad
\mathcal{C}^+ = C\,\Pi_+\approx \left(
\begin{array}{lll}
C & -x_- & -ix_- \\
-x_- & C & 0 \\
-ix_- & 0 & C
\end{array}\right)\,.\;
\end{equation}
Here $\approx$ means after using Eq. \eqref{eqNR1}, and we put
$x_-=x_1-ix_2$.

To prove that this operators are symmetries,
 notice first that \eqref{DC} are elements of the universal enveloping
 algebra of the Galilei algebra, namely polynomials in the Galilei
 algebra generators (boosts and translations). Now, we write Eqns
 \eqref{eqNR1}, \eqref{KGSch} and (\ref{eqNR2}) symbolically as
 $ \mathfrak{D} \Phi = 0 $, where $\mathfrak{D}$ is the respective differential operator.

 Consider now two operators ${\cal A}$  and  ${\cal B}$
 such that $[\,\mathfrak{D},\,{\cal A}]=[\,\mathfrak{D},\,{\cal B}\,]=0$.
 They both preserve the space of solutions of $ \mathfrak{D} \Phi=0$,
 and can be treated therefore as symmetry generators.
 Then it is straightforward to show that
 the product of two such symmetry generators
 ${\cal A}{\cal B}$, is also a symmetry
 $[\,\mathfrak{D},\,{\cal A}{\cal B}\,]=0$. Choosing, in particular,
 ${\cal A}$ and ${\cal B}$ to be Galilei boost or
  momentum generators,  it follows that $\mathcal{C}$ and $\mathcal{D}$,  constructed
of them according to (\ref{DC}), are also (explicitly
t-dependent) symmetries.

 The system of equation \eqref{eqNR1}, \eqref{KGSch} and \eqref{NRDJT}
 is therefore Schr\"odinger symmetric.
The same arguments explain the origin of the
helicity supercharges (\ref{helicity}) introduced above.

Now the superGalilei symmetry combines with the conformal extension,
\beq
    \IS=\frac{1}{2m}\,\vec{\mfQ}\cdot \overrightarrow{\cK}\,,
    \qquad
    \IS^\star=\frac{1}{2m}\,\vec{\mfQ}\times
    \overrightarrow{\cK}\,,
\label{superexp}
\eeq 
related by $\IS^\star=iR\IS$, are both
supercharges for the non-relativistic system. They are both
``square roots'' of expansions, \beq\barray{c} \{ \IS, \,
\IS \}=\{ \IS^\star, \, \IS^\star \}=2\cC,\qquad \{ \IS, \,
\IS^\star \}=0\,. \earray \eeq
Moreover,
\begin{equation}
\{\IQ,\,\IS \}=\{\IQ^\star,\,\IS^\star \}=-\mathcal{D}\,,
\qquad  \{\IQ^\star ,\,\IS \}=-\{\IQ,\,\IS^\star \}=\mathcal{Y} \,,
\end{equation}
where
$\mathcal{Y}=\frac{1}{m}\,\overrightarrow{\mathcal{K}}\times
\vP.$ On shell, $\mathcal{Y}$ is just ${\rm
J}-\hat{\alpha}$, cf. \cite{SchrSusy,AnaP}. Note for further
reference that the conserved quantities (\ref{superexp})
are obtained  by the commuting the generator of the special
conformal transformations, $\mathcal{C}$, with the
supercharges $\IQ$, $\IQ^\star$, see below.

All these generators close, at last, into an $N=2$
superSchr\"odinger algebra \cite{SchrSusy},
$\mathfrak{Ssch}$, that includes the supercharges $\mfQ_i$,
$\IQ$, $\IQ^\star$,  $\IS$ and $\IS^\star $, with
additional commutation relations
\begin{eqnarray}
\begin{array}{llll}
[\cD,\,\IQ]=-i\IQ \,, & [\cC,\,\IQ]=i\IS \,,& [\mathcal{Y},\,\IQ]=i\IQ^\star\,,&
\\[6pt]
[\cD,\,\IQ^\star]=-i\IQ^\star \,,
& [\cC,\,\IQ^\star]=i\IS^\star \,,& [\mathcal{Y},\,\IQ^\star]=-i\IQ\,,&
\\[6pt]
[\cH,\,\IS]= -i\IQ\,,& [\cD,\,\IS]= i \IS \,,& [\cC,\,\IS]= 0 \,,
& [\mathcal{Y},\,\IS]= i \IS^\star \,,
\\[6pt]
[\cH,\,\IS^\star]=-i\IQ^\star \,,
& [\cD,\,\IS^\star]= i \IS^\star \,,& [\cC,\,\IS^\star]= 0\,,
& [\mathcal{Y},\,\IS^\star]= -i\IS \,,
\\[6pt]
[\IS,\,\cK_i ]=0 \,,& [\IS,\,P_i ]=\frac{i}{2}\mfQ_i \,, & [\IS,\,\cH]=i\IQ\,,
\\[6pt]
[\IS^\star,\,\cK_i ]=0 \,,& [\IS^\star,\,P_i ]=-
\frac{i}{2}\epsilon_{ij}\mfQ_j \,, & [\IS^\star,\,\cH]=i\IQ^\star\,.
\end{array}
\end{eqnarray}

\section{The relativistic origin of (super)Schr\"odinger symmetry}

It is well-known that while the Galilei symmetry
is obtained
from the Poincar\'e symmetry by contraction,
Schr\"odinger symmetry, its conformal extension,
can not be derived in such a way \cite{Barut}.
Below we show, however, that the latter, and in fact
superGalilei symmetry, can be
obtained from a relativistic theory, --- but one has to start with a larger structure.

 Consider all
operators which are quadratic in the generators of the
superPoincar\'e algebra. Their commutators
with the superPoincar\'e generators are again
quadratic. The commutators of the quadratic
operators between themselves give rise, however, to the
operators which are cubic in the superPoincar\'e
generators. Continuing this procedure, we end with the
universal enveloping algebra of the superPoincar\'e
algebra.

Restricting ourselves to a certain subset of the
quadratic operators, apply the similarity
transformation (\ref{similarity}) to the commutators of
these operators between themselves, and with the generators
of the superPoincar\'e algebra; then divide both sides of
these commutation relations by appropriate powers of
velocity of light, $c$, and take, finally, the limit $c\rightarrow
\infty$. This procedure can yield a closed Lie superalgebra structure.

To identify an appropriate quadratic subset of the
universal enveloping algebra, we note that the
non-relativistic symmetry generators (\ref{DC}),
(\ref{helicity}) and (\ref{superexp}) can be identified as
the non-relativistic limits of
\begin{equation}\label{tilDK}
    \tilde{\mathcal{D}}=\frac{1}{2m}\epsilon_{ij}(P_i\mathcal{M}_j +
    \mathcal{M}_jP_i)\,,\quad
    \tilde{\mathcal{C}}=\mathcal{M}_i\mathcal{M}_i\,,
\end{equation}
\begin{equation}\label{tildSQ}
    \quad\tilde{\IQ}=\frac{1}{2\sqrt{2}\,m}
    \mathcal{Q}_{\underline{1}}P_-
    +\mathcal{Q}_{\underline{2}}P_+\,,\quad
    \tilde{\IS}=\frac{i}{2\sqrt{2}\,m}
    \left(\mathcal{Q}_{\underline{2}}\mathcal{M}_+
    -\mathcal{Q}_{\underline{1}}\mathcal{M}_-\right)\,,
\end{equation}
and of $\tilde{\IQ}^\star=iR\tilde{\IQ}$ and
$\tilde{\IS}^\star=iR\tilde{\IS}$,
\begin{equation}
    \mathcal{D}=\lim_{c\rightarrow\infty}\tilde{\mathcal{D}}'/c\,,\quad
    \mathcal{C}=\lim_{c\rightarrow \infty}
    \tilde{\mathcal{C}}'/c^2\,,
    \quad\IQ=\lim_{c\rightarrow\infty} \tilde{\IQ}'/c\,,
    \quad\IS=\lim_{c\rightarrow\infty} \tilde{\IS}'/c^2\,.
    \label{limtilD}
\end{equation}

 If we take now, for instance, the commutator of the
$\tilde{\mathcal{D}}$ with $P_i$, we get a new element of
the universal enveloping algebra of the superPoincar\'e
algebra, $[\tilde{\mathcal{D}},P_i]=-\frac{i}{m}P_iP_0$. On
account of the definition (\ref{limtilD}) and of the first relation
from (\ref{Pprime}), this reduces, after applying the
similarity transformation and taking the
non-relativistic limit, to one of the Lie algebraic
relations from (\ref{CDH}), $[\mathcal{D},P_i]=-iP_i$. One
can check then that in a similar way all the rest of the
(anti)commutation relations of the superSchr\"odinger
algebra can be reproduced proceeding from (\ref{tilDK}),
(\ref{tildSQ}) and (\ref{limtilD}).

In conclusion, Schr\"odinger supersymmetry is inherited
from its relativistic predecessor, but this requires the
extension of the superPoincar\'e algebra
 by  certain elements of its universal enveloping algebra,
 which, in the nonrelativistic
 limit, become genuine space-time transformations.

\section{Discussion}

Contraction from the Poincar\'e algebra  yields the
Galilei algebra.
Extending the contraction to the whole superPoincar\'e
algebra only yields \emph{some, but  not all}, of the nonrelativistic symmetries.

Those which are not obtained emerge from higher order
tensor products of the Galilei generators and the
supercharges $\mfQ_i$ in \eqref{Q1NR}-\eqref{Q2NR}. These
products form indeed a finite subset of the universal
enveloping algebra of the Galilei algebra \eqref{Gal}
extended with the supercharges $\mfQ_i$. The latter is
endowed with the supercommutator product, cf.
\cite{COP,MV}. It follows that the new generators descend
from some of the generators of the universal enveloping of
 the superPoincar\'e algebra \eqref{Poinc}-\eqref{SuperPoi1}
  when the nonrelativistic limit is taken.
 In contrast, the relativistic counterpart of the
 new nonrelativistic symmetries would not close into a
 finite dimensional superextension of Poincar\'e,
 but generates instead its whole universal enveloping algebra.

We mention that this approach allowed to show  that
a free scalar nonrelativistic particle in $d$-spatial dimensions exhibits an
$Sp(2d)$ symmetry, which extends its well known conformal
$Sl(2,\mathbb{R})\approx Sp(2)$ symmetry \cite{MV}.

Note that
 the contraction endows the
 higher-order operators \eqref{tilDK} and \eqref{tildSQ} of the
universal enveloping superPoincar\'e algebra,
 with a
clear-cut geometrical meaning: dilations and expansion
generators are genuine space-time transformations, with the
supercharges becoming the square roots of the time
translations and expansions, respectively.

Our considerations here are based on a particular
representation of the
 deformed Heisenberg algebra
\cite{DHA}, that carries a suitable irreducible
representation of $\mathfrak{osp}(1|2)$. It is this
representation that is promoted to an irreducible
representation of the superPoincar\'e symmetry of the
Dirac--DJT system. The advantage of this approach is that
it allows us to work with physical fields only and  to
identify the supersymmetries of the corresponding
nonrelativistic limit. It has also a universal character,
since by taking other representations of the deformed
algebra, one can describe any (including $N$-extended and
anyonic) representations of the superPoincar\'e algebra
\cite{HPV3}. However, unlike in the superfield formulation
\cite{Sieg,Schonf}, the supersymmetry algebra is closed
here only on-shell. This is a price we pay for a
minimality of the supermultiplet  involving no auxiliary fields.

Let us note that the supersymmetry studied here has been, since, extended to anyons \cite{Horvathy:2010we}.


\vskip 0.4cm\noindent {\bf Acknowledgements}. MSP is
indebted to the {\it Laboratoire de Math\'ematiques et de
Physique Th\'eorique} of Tours University, and PAH is
indebted to the {\it Departamento de F\'{\i}sica,
Universidad de Santiago de Chile}, respectively, for
hospitality.
 Partial support by the FONDECYT (Chile)
under the grant 1095027 and by DICYT (USACH), and by
Spanish Ministerio de Educaci\'on  under Project
SAB2009-0181 (sabbatical grant of MSP), is acknowledged. MV
has been supported by CNRS postdoctoral grant (contract
number 87366). We are grateful to S. Deser, R. Jackiw and J. Lukierski for correspondence. We thank to P.M. Zhang for careful reading of our draft.


\end{document}